\documentclass[aps,pra,reprint,10pt,nosuperscriptaddress,amssymb,amsfonts,longbibliography,nofootinbib]{revtex4-1}

\usepackage{amsmath}
\usepackage{graphicx}
\usepackage{dcolumn}
\usepackage{bbm}
\usepackage{subfigure}
\usepackage{amssymb}
\usepackage{wasysym}
\usepackage{MnSymbol}
\usepackage{color}
\usepackage{mathrsfs}
\usepackage[normalem]{ulem}

\stepcounter{secnumdepth}
\stepcounter{tocdepth}

\begin{document}

\title{Electric Circuit Realizations of Fracton Physics}

\author{Michael Pretko}
\affiliation{Department of Physics and Center for Theory of Quantum Matter, University of Colorado, Boulder, CO 80309, USA}

\date{\today}
\begin{abstract}
We design a set of classical macroscopic electric circuits in which charge exhibits the mobility restrictions of fracton quasiparticles.  The crucial ingredient in these circuits is a transformer, which induces currents between pairs of adjacent wires.  For an appropriately designed geometry, this induction serves to enforce conservation of dipole moment.  We show that a network of capacitors connected via ideal transformers will forever remember the dipole moment of its initial charge configuration.  Relaxation of the dipole moment in realistic systems can only occur via flux leakage in the transformers, which will lead to violations of fracton physics at the longest times.  We propose a concrete diagnostic for these ``fractolectric" circuits in the form of their characteristic equilibrium charge configurations, which we verify using simple circuit simulation software.  These circuits not only provide an experimental testing ground for fracton physics, but also serve as DC filters.  We outline extensions of these ideas to circuits featuring other types of higher moment conservation laws, as well as to higher-dimensional circuits which act as fracton ``current-ice."  While our focus is on classical circuits, we discuss how these ideas can be straightforwardly extended to realize quantized fractons in superconducting circuits.
\end{abstract}

\maketitle

\emph{Introduction}.    Advances in the study of quantum phases of matter over the past several decades have demonstrated the existence of numerous phenomena, such as topologically protected edge modes and fractionally-charged quasiparticles, which appear quite exotic from the perspective of classical physics.  However, some of these phenomena also have clear analogues in simpler classical systems.  For example, the robust edge modes seen in topological insulators can be found in both mechanical systems \cite{kanelub,vit,nash,meeu,zero} and ordinary electric circuits \cite{circ1,circ2,topol,zhao,hofmann,nonherm,helbig,active}.  In the latter context, a special class of AC circuits with topological admittance bands featuring robust boundary modes have been both theoretically designed and practically implemented.  Similar work has also taken place on realizing the corner modes associated with certain higher order topological insulators \cite{higher,ezawa,serra}.

While edge modes can be realized in classical systems in straightforward fashion, fractionalized quasiparticles represent a more significant challenge.  The most common types of fractionalized quasiparticles are characterized by fractionalized charge and braiding statistics, which do not have obvious analogues in mechanical or electrical systems (though recent progress has been made in this direction \cite{braid1,braid2}).  In contrast, recent years have uncovered the existence of a striking new type of fractionalized quasiparticle, the ``fracton," characterized by its unusual restricted mobility \cite{chamon,haah,fracton1,fracton2,sub,review}.  Specifically, an isolated fracton is strictly immobile, while certain bound states of fractons are free to move around the system.  This mobility restriction is naturally encoded in the higher moment conservation laws of such systems, such as conservation of dipole moment \cite{sub,higgs1,higgs2}.  Fractons are notable both for their potential applications to quantum information storage \cite{haah,bravyi,terhal,brown}, as well as their prevalence across numerous domains of physics, including spin liquids \cite{slaglenn,field,layer1,layer2,generic,aniso,cage,fusion,pyro,cheng,stab,
twist1,twist2,twist3,fuji}, elasticity \cite{elasticity,gromov,elasticity3d,kumar,z3,prb,leomike}, localization \cite{chamon,glassy,localization}, hole-doped antiferromagnets \cite{polaron}, gravity \cite{mach,holo1,holo2}, Majorana systems \cite{fracton1,lego,majfol}, and deconfined quantum criticality \cite{deconfined}.

While this type of fractionalization is in some ways more exotic than fractionalized statistics, it also has a much clearer path towards realization in classical systems.  The key ingredient in a classical realization should be a mechanism for enforcing conservation of higher charge moments without fine-tuning.  In this work, we provide precisely such a mechanism in the context of ordinary electric circuits.  Specifically, we show how transformers can be utilized in circuits to enforce conservation of dipole moment.  We then design a set of circuits featuring capacitors and transformers, which we term ``fractolectric circuits," in which electric charge inherits the mobility restrictions of fractons.  The dipole conservation in these circuits is quite robust, insensitive to internal resistances in the circuit.  The only effect which violates this constraint is transformer flux leakage, which causes the dipole to relax at the longest times in realistic systems.

We propose several concrete ways to characterize fractolectric circuits.  For example, the steady-state charge distribution of such a circuit has a characteristic linear form, as we describe in detail below.  We verify this prediction using the circuit simulator CircuitLab, in which we design one-dimensional circuits explicitly exhibiting dipole conservation.  We also argue these circuits act as perfect DC filters.  Such circuits could be readily built in table-top experiments and would provide a new platform for testing the physics of fractons.  We conclude by outlining a procedure for systematically imposing higher moment conservation laws beyond dipole moment into electric circuits, including those leading to other types of subdimensional particles besides fractons.  We also consider higher-dimensional circuits which act as fracton ``current-ice," exhibiting pinch-point singularities in their current-current correlations.  While we focus on macroscopic classical circuits, we briefly discuss extensions to superconducting quantum circuits realizing quantized fractons.

\begin{figure}[t!]
 \centering
 \includegraphics[scale=0.40]{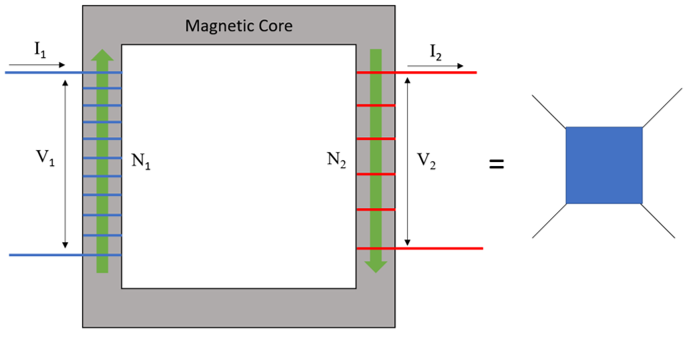}
 \caption{A transformer functions via a central magnetic core which ideally allows all magnetic flux (green arrow) generated by an input current to pass through a secondary coil.  The ratio of voltages equals the ratio of number of windings of the two coils, $i.e.$ $V_1/V_2 = N_1/N_2$.  We abstractly represent a transformer as a box with four contact points.}
 \label{fig:trans}
\end{figure}

\emph{Circuit Design}.    A transformer is a component of electric circuits which is primarily used for transforming alternating current from one voltage in one wire to a different voltage in another wire.  The key physical principle at work in a transformer is electromagnetic induction, whereby an alternating current in the primary wire induces an alternating current in the secondary wire.  A simple physical implementation of a transformer involves both wires being coiled around a magnetic core, in general with different numbers of windings (see Figure \ref{fig:trans}).  When the core material is chosen to have extremely high magnetic permeability, essentially all of the magnetic flux generated by a current in the primary wire will pass through the coil of the secondary wire.  When the primary current is alternating, the changing flux will induce an alternating current in the secondary wire at the same frequency.

While the primary and secondary wires carry alternating currents at the same frequency, they are generically at different voltages.  Specifically, the ratio of the primary voltage to the secondary voltage is simply the ratio of the number of times their respective wires are wound around the central core, $V_1/V_2 = N_1/N_2$, as a simple consequence of Faraday's law.  Importantly, these ratios can be negative if the coils are wound around the core in opposite orientations.  For the purpose of realizing fracton physics, we will choose transformers designed such that this ratio is $-1$, $i.e.$ both coils have the same number of windings, but with opposite orientation, so that the secondary voltage is simply inverted with respect to the primary.  These voltages can then be related to the currents passing through the two wires, or more specifically, the time derivative of current.  The voltage across each coil can be written as:
\begin{equation}
V_{1/2} = - L\partial_t I_{1/2} - M \partial_t I_{2/1}
\end{equation}
where $L$ is the self-inductance of each of the two coils (which is the same for both, since the wires are taken to be identical up to orientation) and $M$ is the mutual inductance of the two coils.  We have neglected any internal resistance, which we discuss below in the context of non-ideal transformers.  By setting $V_1 = -V_2$, we can then conclude that:
\begin{equation}
\frac{dI_1}{dt} = -\frac{dI_2}{dt}
\end{equation}
If we Fourier transform to the frequency domain, then away from $\omega = 0$ ($i.e.$ the DC component), we can conclude that $\tilde{I}_1(\omega) = -\tilde{I}_2(\omega)$.  Alternatively, if we stipulate that $I_1$ and $I_2$ are equal and opposite at $t=0$, we can conclude that they remain equal and opposite for all times:
\begin{equation}
I_1(t) = -I_2(t)
\label{dc}
\end{equation}
We therefore conclude that, up to a constant DC offset, the two currents remain equal and opposite at all times.  We could also have independently reached the same conclusion based on energy conservation.  Neglecting internal resistance, we can match the power input and output of the two wires, yielding $I_1V_1 = I_2V_2$.  If the voltages are equal and opposite, then so too are the currents.

We now have a circuit element which enforces a perfect ``drag" effect, in the sense that current in one wire leads to equal and opposite current in some nearby wire.  This physics is highly reminiscent of the behavior of fractons, for which motion of a charge is necessarily accompanied by opposing motion of nearby charges in such a way as to preserve the overall dipole moment.  Indeed, by storing charge in an appropriate geometry, we can use these transformers to construct a circuit which explicitly enforces conservation of dipole moment.  For ease of notation, we represent transformers simply as an abstract box with four connection points, as in Figure \ref{fig:trans}.  (Note that there is no ambiguity in this notation for the special case of $N_1/N_2 = -1$.)  We also use circles to represent one end of a capacitor (with the other end implicitly grounded).

\begin{figure}[t!]
 \centering
 \includegraphics[scale=0.42]{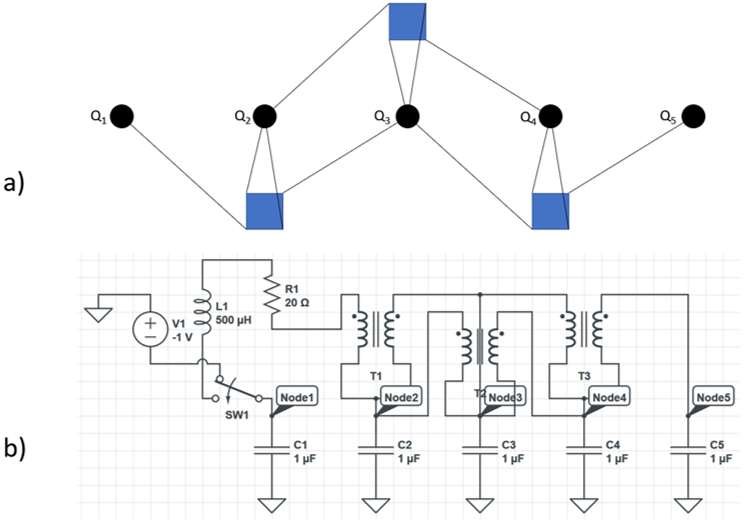}
 \caption{a) Schematic of a dipole-conserving fractolectric circuit, with black circles representing capacitors and blue squares representing transformers.  b) Implementation of a fractolectric circuit in CircuitLab.  Each capacitor has $C = 1\,\mu$F, the self and mutual inductances of coils in the transformers are 10 H, and external resistance and inductance of 20 $\Omega$ and 0.5 H have been added to regulate the circuit.}
 \label{fig:chain}
\end{figure}

We consider a lattice of such capacitors, which carry all charge in the system.  We focus on a one-dimensional chain, though the principal extends to arbitrary dimension without difficulty.  We connect the capacitors of the chain using transformers, designed so that current through any link between neighboring capacitors induces an opposite current in a nearby link.  An example schematic of this type of circuit is shown in Figure \ref{fig:chain}a.  Note that two different current paths are available between any two neighboring capacitors in the bulk of the chain, corresponding to the possibility of inducing an opposing current in either the leftward pair or rightward pair of capacitors.  This circuit explicitly exhibits conservation of dipole moment, by design.  One way to verify this is by deriving the generalized continuity equation of the circuit.  If a transformer at location $x_n$ carries current $I(x_n)$ ($i.e.$ $I(x_n)$ in one wire and $-I(x_n)$ in the other), it is straightforward to verify that, in the bulk of the system, the charge $Q(x_n)$ obeys the following relation:
\begin{equation}
\partial_tQ + \partial_x^2I = 0
\label{cont}
\end{equation}
where the spatial derivatives should be interpreted as lattice differences.  We then conclude that the change in dipole moment, $\partial_t (\sum_n Q_n x_n) = -\sum_n x_n\partial_x^2 I_n$, is a boundary term, which vanishes with the boundary conditions chosen in Figure \ref{fig:chain}.

\emph{Diagnostics}.    Given that the circuit indicated by Figure \ref{fig:chain}a should exhibit conservation of dipole moment, what physical observable can we examine to test this?  One particularly simple metric for dipole conservation is the steady-state charge distribution.  We assume that the circuit contains a small internal resistance which eventually causes currents to relax and the charge to reach a steady state.  (Importantly, dipole conservation, which follows from equality of flux on the two sides of a transformer, is not affected by equal resistances added to both sides.)  Consider a chain of identical capacitors initialized with some non-uniform charge distribution.  In the absence of dipole conservation, the charge would eventually spread out evenly in order to minimize energy, such that each capacitor carried equal charge.  In the presence of dipole conservation, however, the chain can no longer relax to the true minimum energy configuration.  Instead, the system will relax to the minimum energy configuration consistent with dipole conservation.  To find this configuration, we minimize the following energy functional:
\begin{equation}
E = \frac{1}{2}C\sum_n Q_n^2 - \mu \sum_n Q_n - \lambda \sum_n x_nQ_n
\end{equation}
where $Q_n$ is the charge on capacitor $n$, $x_n$ is its position, $C$ is the capacitance of each capacitor (assumed uniform), and $\mu$ and $\lambda$ are Lagrange multipliers which we will use to enforce particular values of charge and dipole moment.  Varying the energy with respect to the $Q_i$, we obtain the minimum energy configuration as:
\begin{equation}
Q_n = \frac{2}{C}(\mu + \lambda x_n)
\end{equation}
where $\mu$ and $\lambda$ are chosen such that this configuration has the same charge and dipole moment as the initial configuration.  Note that, for a given total charge, $\lambda$ can be zero for only a single value of the dipole moment.  Generically, $\lambda$ is nonzero.  In other words, the steady state charge distribution is a linear function of $x$, instead of a uniform distribution.

\begin{figure}[t!]
 \centering
 \includegraphics[scale=0.38]{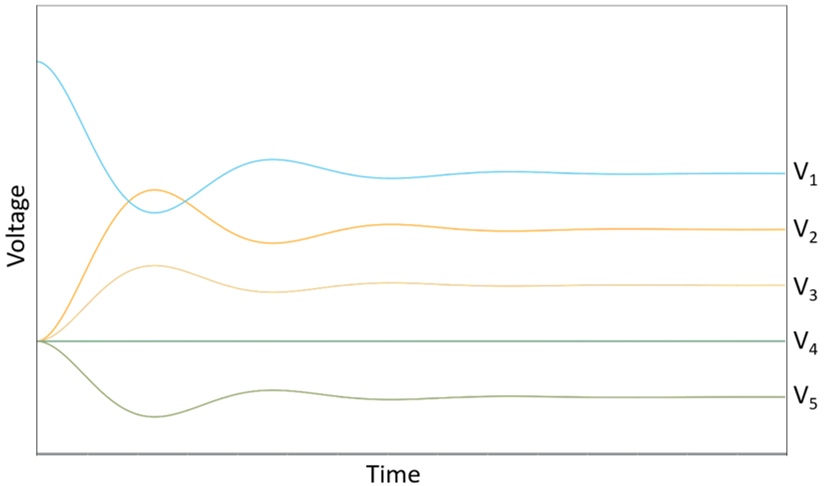}
 \caption{Voltage across each capacitor (from -0.4 to 1.2 V) as a function of time (from 0 to 0.3 ms), after initializing the system with 1 V across the leftmost capacitor.  At late times, the charge is a linear function of position, as predicted.}
 \label{fig:volt}
\end{figure}

We can test this prediction by directly simulating a fractolectric circuit using simple circuit simulation software.  In Figure \ref{fig:chain}b, we display a fractolectric circuit built using CircuitLab which will allow us to put these ideas to the test.  We can easily check the steady state charge distribution by reading off the voltage across each capacitor.  We consider initializing the system with charge $1$ on the leftmost capacitor, then we let the system relax to a steady state.  In Figure \ref{fig:volt}, we plot the voltage across each capacitor as a function of time.  After some initial oscillations, the system reaches a steady state in which charge behaves as a linear function of position, just as predicted, serving as a clear indication of conservation of dipole moment.

It is also useful to consider how such a circuit responds to an externally applied voltage.  A direct current can easily pass through the circuit, which only places restrictions on changing currents.  For an applied voltage difference $V$ across the two terminal points of the circuit, the system will exhibit a current $I = V/R_{eff}$, where $R_{eff}$ is the effective resistance generated by all internal components of the circuit.  For a purely alternating applied voltage, however, the conservation of dipole moment will not allow any net flow of charge from one end of the circuit to the other, acting as an infinite impedance.  More generally, for an applied voltage $V(t)$, the fractolectric circuit will only allow passage of the DC component, $\tilde{V}(\omega = 0)$.  Thus, neglecting small losses due to flux leakage, the fractolectric circuit acts as a perfect DC filter.

\begin{figure}[t!]
 \centering
 \includegraphics[scale=0.5]{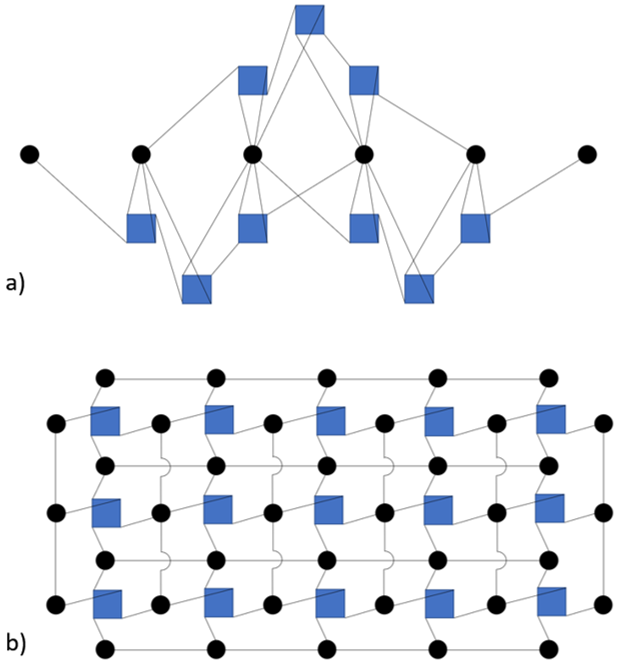}
 \caption{a)  Multiple layers of transformers can be used to construct a circuit which conserves both dipole and quadrupole moment.  b)  Schematic of a two-dimensional circuit featuring one-dimensional particles.}
 \label{fig:quad}
\end{figure}

\emph{Extensions}.    So far, we have considered classical circuits which use transformers to implement conservation of dipole moment.  However, there are various ways in which this idea can be extended.  One natural question to ask is whether this method can be used to implement conservation of even higher multipoles, such as quadrupole moment.  We answer this question in the affirmative by providing an explicit example in Figure \ref{fig:quad}a, which presents a circuit conserving both dipole and quadrupole moment.  The circuit requires a more complex arrangement of transformers, with a hierarchical layered structure.  Each cluster of three transformers in this circuit constrains the currents between four neighboring sites, allowing for currents only in a characteristic pattern conserving the quadrupole moment.  This method can be extended up to conserve arbitrarily high charge moments by adding further layers of transformers.

It is also interesting to consider extensions to higher dimensions.  A higher-dimensional lattice of capacitors can be made to conserve dipole moment in fairly similar fashion to our one-dimensional example, though the circuit diagrams rapidly become cumbersome to draw.  In higher dimensions, we can have transformers which tie together the current in two $i$-directed wires separated in the $j$ direction, or equivalently two $j$-directed wires separated in the $i$ direction.  In other words, the current through each transformer can be regarded as a tensor, $I_{ij}(x_n)$.  The generalized continuity condition of Equation \ref{cont} will then become:
\begin{equation}
\partial_t Q + \partial_i\partial_jI^{ij} = 0
\end{equation}
A particularly interesting limit to examine is when $C\rightarrow 0$, such that no charge is stored anywhere in the circuit and we require $\partial_i\partial_jI^{ij}=0$.  We expect that such a system will generically have an energy functional of the form $E = \sum_n I^{ij}I_{ij}$, where $I_{ij}$ is subject to the constraint $\partial_i\partial_jI^{ij}=0$.  The current-current correlations in this system, $\langle I_{ij}(x)I_{k\ell}(y)\rangle$, then ought to exhibit the ``pinch-point" singularities characteristics of $U(1)$ fracton systems \cite{pinch}.  In this way, these circuits serve as a fracton ``current-ice," in analogy with the spin-ices found in frustrated magnets.

In addition to fractons, higher-dimensional systems can host particles which have mobility restrictions only along certain directions.  For example, in certain ``vector charge" models, an individual charge is restricted to move in a single direction, while perpendicular motion can only occur in bound states.  Realizing this type of physics in circuits is not much more difficult than the simpler fracton case.  Motivated by microscopic models for one-dimensional particles, where charges typically live on links of a lattice, we design a circuit with capacitors on each link of a square lattice, as depicted in Figure \ref{fig:quad}b.  Current can flow normally between capacitors along a fixed line, while motion perpendicular to each line is governed by a set of transformers on all plaquettes of the square lattice.  By grouping the charge on an $x$- and $y$-directed link touching site $n$ into a vector $(Q_x,Q_y)_n$, it can readily be verified that the circuit of Figure \ref{fig:quad}b exhibits conservation of the angular charge moment, $\sum_n \epsilon^{ij}Q_ix_j$, enforcing the one-dimensional nature of charge.  By introducing more complicated geometries, we can further generalize this logic to endow charge with any desired type of mobility.

\emph{Quantum Circuits}.    While we have so far focused on classical circuits, our ideas can also be naturally implemented in superconducting quantum circuits, which provides at least three significant technical advantages.  First, due to the direct relationship between the current in a superconducting loop and the magnetic flux through that loop, the use of superconducting wires would enforce Equation \ref{dc} even for direct currents, not just alternating currents.  Second, a superconducting flux transformer \cite{flux1,flux2,flux3,flux4} could be used to achieve direct transfer of flux from one coil to the other, eliminating violations of fracton behavior arising from flux leakage out of an imperfect core material.  Finally, in contrast to the macroscopic charges carried by classical capacitors, a quantum circuit could store quantized charges via the use of quantum dots.  All of these features add up to a more robust realization of fractons in quantum circuits, as compared with their classical counterparts.  Using these techniques, a quantum circuit consisting of only quantum dots and superconducting wires could be used to realize the physics of discrete charges exhibiting perfect fracton behavior, even in the DC limit.

\emph{Conclusions}.    In this work, we have established a design for realizing the constrained dynamics of fractons in ordinary classical macroscopic electric circuits.  These circuits rely on the induction physics of transformers to naturally enforce conservation of dipole moment, allowing for a simple classical realization of the fracton phenomenon.  We have shown that the charge distribution on a network of capacitors connected through appropriate transformers will be forced to remember its initial dipole moment, instead of relaxing to its true minimum energy configuration.  We have proposed various probes of the fractonic nature of these circuits, which we have verified using simple circuit simulation software.  Finally, we have outlined extensions of this circuit design which conserve other higher multipole moments, such as those leading to subdimensional behavior.  We have also considered higher-dimensional fractolectric circuits which behave as a fracton current-ice, exhibiting characteristic pinch-point singularities.  Our work opens the door for simple table-top experiments on fracton physics.  We have also discussed the extension of these ideas to superconducting quantum circuits, which allow for perfect realization of quantized fractons.

\begin{acknowledgments}
I would particularly like to thank Sid Parameswaran for several insightful discussions and for his hospitality at Oxford University, where much of this work was completed.  I am also grateful to Shriya Pai, Drew Potter, Nicholas Watkins, and Yang-Zhi Chou for useful conversations.
\end{acknowledgments}


\begin{thebibliography}{9}

\bibitem{kanelub}
C. L. Kane and T. C. Lubensky, \emph{Topological boundary modes in isostatic lattices}.  Nature Physics 10, 39 (2014), arXiv:1308.0554v2

\bibitem{vit}
B. G.-g. Chen, N. Upadhyaya, and V. Vitelli, \emph{Nonlinear conduction via solitons in a topological mechanical insulator}.  PNAS, vol. 111 no. 36 13004-13009 (2014), arXiv:1404.2263v2

\bibitem{nash}
L. M. Nash, D. Kleckner, A. Read, V. Vitelli, A. M. Turner, and W. T. M. Irvine, \emph{Topological mechanics of gyroscopic metamaterials}.  Proc. Natl. Acad. Sci. 112 (2015), 14495-14500, arXiv:1504.03362v2

\bibitem{meeu}
A. S. Meeussen, J. Paulose, and V. Vitelli, \emph{Geared topological metamaterials with tunable mechanical stability}.  Phys. Rev. X 6, 041029 (2016), arXiv:1602.08769v2

\bibitem{zero}
N. Lera and J. V. Alvarez, \emph{Mechanical topological insulator in zero dimensions}.  Phys. Rev. B 97, 134118 (2018)

\bibitem{circ1}
N. Jia, C. Owens, A. Sommer, D. Schuster, and J. Simon, \emph{Time reversal invariant topologically insulating circuits}.  Phys. Rev. X 5, 021031 (2015), arXiv:1309.0878v2

\bibitem{circ2}
V. V. Albert, L. I. Glazman, and L. Jiang, \emph{Topological properties of linear circuit lattices}.  Phys. Rev. Lett. 114, 173902 (2015), arXiv:1410.1243v3

\bibitem{topol}
C. H. Lee et al., \emph{Topolectrical circuits}.  Communications Physics, Vol 1, 39 (2018), arXiv:1705.01077v3

\bibitem{zhao}
E. Zhao, \emph{Topological circuits of inductors and capacitors}.  Annals of Physics 399, 289-313 (2018), arXiv:1810.10318v2

\bibitem{hofmann}
T. Hofmann, T. Helbig, C. H. Lee, M. Greiter, and R. Thomale, \emph{Chiral voltage propagation and calibration in a topolectrical Chern circuit}.  Phys. Rev. Lett. 122, 247702 (2019), arXiv:1809.08687v2

\bibitem{nonherm}
M. Ezawa, \emph{Electric circuits for non-Hermitian Chern insulators}.  arXiv:1904.03823v2 (2019)

\bibitem{helbig}
T. Helbig, T. Hofmann, C. H. Lee, R. Thomale, S. Imhof, L. W. Molenkamp, and T. Kiessling, \emph{Band structure engineering and reconstruction in electric circuit networks}.  Phys. Rev. B 99, 161114(R) (2019), arXiv:1807.09555

\bibitem{active}
T. Kotwal, H. Ronellenfitsch, F. Moseley, and J. Dunkel, \emph{Active topolectrical circuits}.  arXiv:1903.10130v2 (2019)

\bibitem{higher}
S. Imhof et al., \emph{Topolectrical circuit realization of topological corner modes}.  Nature Physics 14, 925–929 (2018), arXiv:1708.03647

\bibitem{ezawa}
M. Ezawa, \emph{Higher-order topological electric circuits and topological corner resonance on the breathing Kagome and pyrochlore lattices}.  Phys. Rev. B 98, 201402(R) (2018), arXiv:1809.08847

\bibitem{serra}
M. Serra-Garcia, R. S\"usstrunk, and S. D. Huber, \emph{Observation of quadrupole transitions and edge mode topology in an LC network}.  Phys. Rev. B 99, 020304 (2019), arXiv:1806.07367

\bibitem{braid1}
M. Ezawa, \emph{Braiding of Majorana corner states in electric circuits and its non-Hermitian generalization}.  Phys. Rev. B 100, 045407 (2019), arXiv:1902.03716v2

\bibitem{braid2}
M. Ezawa, \emph{Non-Abelian braiding of Majorana-like edge states and scalable topological quantum computations in electric circuits}.  arXiv:1907.06911 (2019)

\bibitem{chamon}
C. Chamon, \emph{Quantum glassiness in strongly correlated clean systems: An example of topological overprotection}.  Phys. Rev. Lett. 94 040402 (2005), arXiv:cond-mat/0404182v2

\bibitem{haah}
J. Haah, \emph{Local stabilizer codes in three dimensions without string logical operators}.  Phys. Rev. A 83, 042330 (2011), arXiv:1101.1962v2

\bibitem{fracton1}
S. Vijay, J. Haah, L. Fu, \emph{A new kind of topological quantum order: A dimensional hierarchy of quasiparticles built from stationary excitations}.  Phys. Rev. B 92, 235136 (2015), arXiv:1505.02576

\bibitem{fracton2}
S. Vijay, J. Haah, L. Fu, \emph{Fracton topological order, generalized lattice gauge theory and duality}.  Phys. Rev. B 94, 235157 (2016), arXiv:1603.04442

\bibitem{sub}
M. Pretko, \emph{Subdimensional particle structure of higher rank U(1) spin liquids}.  Phys. Rev. B 95, 115139 (2017), arXiv:1604.05329v3

\bibitem{review}
R. M. Nandkishore and M. Hermele, \emph{Fractons}.  Annual Review of Condensed Matter Physics, 10, 295-313 (2019), arXiv:1803.11196

\bibitem{higgs1}
H. Ma, M. Hermele, and X. Chen, \emph{Fracton topological order from Higgs and partial confinement mechanisms of rank-two gauge theory}.  Phys. Rev. B 98, 035111 (2018), arXiv:1802.10108v2

\bibitem{higgs2}
D. Bulmash and M. Barkeshli, \emph{The Higgs mechanism in higher-rank symmetric $U(1)$ gauge theories}.  Phys. Rev. B 97, 235112 (2018), arXiv:1802.10099v2

\bibitem{bravyi}
S. Bravyi and J. Haah, \emph{Quantum self-correction in the 3d cubic code model}. Phys. Rev. Lett. 111, 200501 (2013), arXiv:1112.3252

\bibitem{terhal}
B. Terhal, \emph{Quantum error correction for quantum memories}.  Rev. Mod. Phys. 87, 307 (2015), arXiv:1302.3428v7

\bibitem{brown}
B. J. Brown and D. J. Williamson, \emph{Parallelized quantum error correction with fracton topological codes}.  arXiv:1901.08061 (2019)

\bibitem{slaglenn}
K. Slagle and Y.-B. Kim, \emph{Fracton topological order from nearest-neighbor two-spin interactions and dualities}.  Phys. Rev. B 96, 165106 (2017), arXiv:1704.03870v2

\bibitem{field}
K. Slagle and Y. B. Kim, \emph{Quantum field theory of X-cube fracton topological order and robust degeneracy from geometry}. Phys. Rev. B 96, 195139 (2017), arXiv:1708.04619v3

\bibitem{layer1}
H. Ma, E. Lake, X. Chen, and M. Hermele, \emph{Fracton topological order via coupled layers}.  Phys. Rev. B 95, 245126 (2017), arXiv:1701.00747v2

\bibitem{layer2}
S. Vijay, \emph{Isotropic layer construction and phase diagram for fracton topological phases}.  arXiv:1701.00762 (2017)

\bibitem{generic}
W. Shirley, K. Slagle, Z. Wang, and X. Chen, \emph{Fracton models on general three-dimensional manifolds}. arXiv:1712.05892 (2017)

\bibitem{aniso}
O. Petrova and N. Regnault, \emph{A simple anisotropic three-dimensional quantum spin liquid with fracton topological order}.  Phys. Rev. B 96, 224429 (2017), arXiv:1709.10094

\bibitem{cage}
A. Prem, S.-J. Huang, H. Song, and M. Hermele, \emph{Cage-net fracton models}.  
Phys. Rev. X 9, 021010 (2019), arXiv:1806.04687v2

\bibitem{fusion}
S. Pai and M. Hermele, \emph{Fracton fusion and statistics}. arXiv:1903.11625 (2019)

\bibitem{pyro}
H. Yan, O. Benton, L. D. C. Jaubert, and N. Shannon, \emph{Rank-2 $U(1)$ spin-liquid on the breathing pyrochlore lattice}.  arXiv:1902.10934 (2019)

\bibitem{cheng}
D. Williamson, Z. Bi, and M. Cheng, \emph{Fractonic matter in symmetry-enriched $U(1)$ gauge theory}.  arXiv:1809.10275 (2018)

\bibitem{stab}
A. Schmitz, \emph{Gauge structure: From stabilizer codes to continuum models}.  arXiv:1809.10151v3 (2018)

\bibitem{twist1}
H. Song, A. Prem, S.-J. Huang, and M. A. Martin-Delgado, \emph{Twisted fracton models in three dimensions}.  Phys. Rev. B 99, 155118 (2019), arXiv:1805.06899v2

\bibitem{twist2}
Y. You, T. Devakul, F. J. Burnell, and S. L. Sondhi, \emph{Symmetric fracton matter: Twisted and enriched}.  arXiv:1805.09800 (2018)

\bibitem{twist3}
W. Shirley, K. Slagle, and X. Chen, \emph{Twisted foliated fracton phases}.  arXiv:1907.09048 (2019)

\bibitem{fuji}
Y. Fuji, \emph{Anisotropic layer construction of anisotropic fracton models}.  arXiv:1908.02257v2 (2019)

\bibitem{elasticity}
M. Pretko and L. Radzihovsky, \emph{Fracton-elasticity duality}. Phys. Rev. Lett. 120, 195301 (2018), arXiv:1711.11044

\bibitem{gromov}
A. Gromov, \emph{Fractional topological elasticity and fracton order}. arXiv:1712.06600 (2017)

\bibitem{elasticity3d}
S. Pai and M. Pretko, \emph{Fractonic line excitations: An inroad from 3d elasticity theory}. Phys. Rev. B 97, 235102 (2018), arXiv:1804.01536

\bibitem{kumar}
A. Kumar and A. C. Potter, \emph{Symmetry enforced fractonicity and 2d quantum crystal melting}. arXiv:1808.05621 (2018)

\bibitem{z3}
Z. Zhai and L. Radzihovsky, \emph{Two-dimensional melting via sine-Gordon duality}.  arXiv:1905.00905 (2019)

\bibitem{prb}
M. Pretko, Z. Zhai, and L. Radzihovsky, \emph{Crystal-to-fracton tensor gauge theory dualities}.  arXiv:1907.12577 (2019)

\bibitem{leomike}
L. Radzihovsky and M. Hermele, \emph{Fractons from vector gauge theory}.  arXiv:1905.06951 (2019)

\bibitem{glassy}
A. Prem, J. Haah, and R. Nandkishore, \emph{Glassy quantum dynamics in translation invariant fracton models}.  Phys. Rev. B 95, 155133 (2017), arXiv:1702.02952

\bibitem{localization}
S. Pai, M. Pretko, and R. Nandkishore, \emph{Localization in fractonic random circuits}.  Phys. Rev. X 9, 021003 (2019), arXiv:1807.09776v3

\bibitem{polaron}
J. Sous and M. Pretko, \emph{Fractons from polarons and hole-doped antiferromagnets: Microscopic realizations}.  arXiv:1904.08424v2 (2019)

\bibitem{mach}
M. Pretko, \emph{Emergent gravity of fractons: Mach's principle revisited}.  Phys. Rev. D 96, 024051 (2017), arXiv:1702.07613v3

\bibitem{holo1}
H. Yan, \emph{Hyperbolic fracton model, subsystem symmetry, and holography}.  Phys. Rev. B 99, 155126 (2019), arXiv:1807.05942v4

\bibitem{holo2}
H. Yan, \emph{Hyperbolic fracton model, subsystem symmetry, and holography II: the dual eight-vertex model}.  arXiv:1906.02305 (2019)

\bibitem{lego}
Y. You and F. von Oppen, \emph{Majorana quantum Lego, a route towards fracton matter}.  arXiv:1812.06091v2 (2018)

\bibitem{majfol}
T. Wang, W. Shirley, and X. Chen, \emph{Foliated fracton order in the Majorana checkerboard model}.  arXiv:1904.01111 (2019)

\bibitem{deconfined}
H. Ma and M. Pretko, \emph{Higher rank deconfined quantum criticality at the Lifshitz transition and the exciton Bose condensate}.  Phys. Rev. B 98, 125105 (2018), arXiv:1803.04980v2

\bibitem{pinch}
A. Prem, S. Vijay, Y.-Z. Chou, M. Pretko, R. M. Nandkishore, \emph{Pinch point singularities of tensor spin liquids}.  Phys. Rev. B 98, 165140 (2018), arXiv:1806.04148

\bibitem{flux1}
D. Duret and P. Karp, \emph{Figure of merit and spatial resolution of superconducting flux transformers}.  Journal of Applied Physics, 56(6):1762-1768 (1984)

\bibitem{flux2}
Y. Shimazu and T. Niizeki, \emph{Characteristics of switchable superconducting flux transformer with dc superconducting quantum interference device}.  Japanese Journal of Applied Physics, Vol. 46, No. 4A, 1478-1481 (2007)

\bibitem{flux3}
H. A. Dyvorne, J. Scola, C. Fermon, J. F. Jacquinot, and M. Pannetier-Lecoeur, \emph{Flux transformers made of commercial high critical temperature superconducting wires}.  Review of Scientific Instruments, 79:025107 (2008)

\bibitem{flux4}
M. F. A. Fernandes, \emph{Squid based cryogenic current comparator for measuring low-intensity antiproton beams}.  Ph.D. Thesis, University of Liverpool (2018)



\end{thebibliography}
\end{document}